\documentclass[12pt]{article}
\usepackage{graphicx,amsfonts,amsmath,amssymb,amsthm,mathrsfs,url}
\usepackage[usenames,dvipsnames]{xcolor}

\title{Avoiding Ultraviolet Divergence by Means of Interior--Boundary Conditions}

\author{
Stefan Teufel\footnote{Mathematisches Institut,
     Eberhard-Karls-Universit\"at, Auf der Morgenstelle 10, 72076
     T\"ubingen, Germany. E-mail:
     stefan.teufel@uni-tuebingen.de},\ 
Roderich Tumulka\footnote{Department of Mathematics, 
     Rutgers University, 110 Frelinghuysen Road, Piscataway, NJ 08854-8019, USA.
     E-mail: tumulka@math.rutgers.edu}
}
\date{May 27, 2015}

\newcommand{\Hilbert}{\mathscr{H}}
\newcommand{\Fock}{\mathscr{F}}
\newcommand{\conf}{\mathcal{Q}}
\newcommand{\Q}{\mathcal{Q}}

\renewcommand{\Im}{\mathrm{Im}}
\newcommand{\RRR}{\mathbb{R}}
\newcommand{\CCC}{\mathbb{C}}
\newcommand{\SSS}{\mathbb{S}}
\newcommand{\sM}{\mathscr{M}}
\newcommand{\scp}[2]{\langle #1|#2 \rangle}

\newcommand{\Laplace}{\Delta}

\newcommand{\orig}{{\mathrm{orig}}}
\newcommand{\free}{{\mathrm{free}}}
\newcommand{\cutoff}{{\mathrm{cutoff}}}

\newcommand{\vq}{\boldsymbol{q}}
\newcommand{\vx}{\boldsymbol{x}}
\newcommand{\vy}{\boldsymbol{y}}
\newcommand{\vomega}{\boldsymbol{\omega}}
\newcommand{\vzero}{\boldsymbol{0}}
\newcommand{\domain}{\mathscr{D}}

\newtheorem{thm}{Theorem}
\newcommand{\be}{\begin{equation}}
\newcommand{\ee}{\end{equation}}
\newcounter{tumulkaremarks}
\begin{document}
\maketitle
\begin{abstract}
We describe here a novel way of defining Hamiltonians for quantum field theories (QFTs), based on the particle--position representation of the state vector and involving a condition on the state vector that we call an ``interior--boundary condition.'' At least for some QFTs (and, we hope, for many), this approach leads to a well-defined, self-adjoint Hamiltonian without the need for an ultraviolet cut-off or renormalization. 

\medskip

\end{abstract}

\section{Introduction}

In quantum field theories (QFTs), the terms in the Hamiltonian governing particle creation and annihilation are usually ultraviolet (UV) divergent. The problem can be circumvented by a UV cut-off, that is, by either discretizing space or attributing a nonzero radius to the electron (or other particles). Another, novel approach \cite{TT15,ibc2a} is outlined here, leading to Hamiltonians that are well defined, involve particle creation and annihilation, treat space as a continuum, and give radius zero to electrons. They are defined in the particle-position representation of Fock space by means of a new kind of boundary condition on the wave function, which we call an \emph{interior-boundary condition} (IBC) because it relates values of the wave function on a boundary of configuration space to values in the interior. Here, the relevant configuration space is that of a variable number of particles, the relevant boundary consists of the collision configurations (i.e., those at which two or more particles meet), and the relevant interior point lies in a sector with fewer particles. 

An IBC is a rather simple condition and provides, as we explain below, a mathematically natural way of implementing particle creation and annihilation at a source of radius zero. It is associated with a Hamiltonian $H_{IBC}$ defined on a domain consisting of functions that satisfy the IBC. For several models that our collaborators Jonas Lampart, Julian Schmidt, and we have studied, we have been able to prove the self-adjointness of $H_{IBC}$; these results ensure that $H_{IBC}$ is free of divergence problems, UV or otherwise. For suitable choice of the IBC, the Hamiltonian also seems physically plausible, for several reasons that we will describe in more detail in Section~\ref{sec:qft}: (i)~$H_{IBC}$ has relevant similarities to the original, UV divergent expression for $H$ that one would guess from physical principles; these similarities make $H_{IBC}$ seem like a natural interpretation of that expression. (ii)~$H_{IBC}$ has properties and consequences that seem physically reasonable. (iii)~In certain models it is possible, after starting from the original
expression for $H$ and introducing a UV cut-off, to obtain a well-defined
limiting Hamiltonian $H_\infty$ by taking a suitable limit of removing the
cut-off; $H_\infty$ is called a renormalized Hamiltonian (see, e.g., \cite{Der03}).
We have found in such cases that $H_{IBC}$ agrees with $H_\infty$ up to addition of a finite constant; this result supports that $H_{IBC}$ is physically reasonable and, conversely, provides an explicit description of $H_\infty$ that was not available so far. 

In this paper, we focus on \emph{non-relativistic} Hamiltonians; there is work in progress \cite{ibc3} about similar constructions with the Dirac operator. Further work on IBCs is forthcoming in \cite{bohmibc,KS15,ibc2b}. A future goal is to formulate quantum electrodynamics (QED) in terms of IBCs, building particularly on the work of Landau and Peierls \cite{LP30} about QED in the particle-position representation. 

This paper is organized as follows. In Section~\ref{sec:ex}, we give a gentle introduction to the idea of an interior--boundary condition by means of a toy quantum theory. In Section~\ref{sec:crea}, we describe a similar IBC and Hamiltonian involving particle creation and annihilation. In Section~\ref{sec:qft}, we describe how this can be applied to QFTs by means of the particle-position representation, and report some results on the rigorous existence of self-adjoint Hamiltonians defined by means of IBCs.

\section{Simple Example of an Interior--Boundary Condition}
\label{sec:ex}

To introduce the concept of an IBC, we start with a toy example, for which we will set up a ``configuration space'' $\Q$, a Hilbert space $\Hilbert=L^2(\Q)$, and a Hamiltonian $H$ on $\Hilbert$.

\subsection{Configuration Space and Hilbert Space}

\begin{figure}[h]
\begin{center}
\includegraphics[width=4cm]{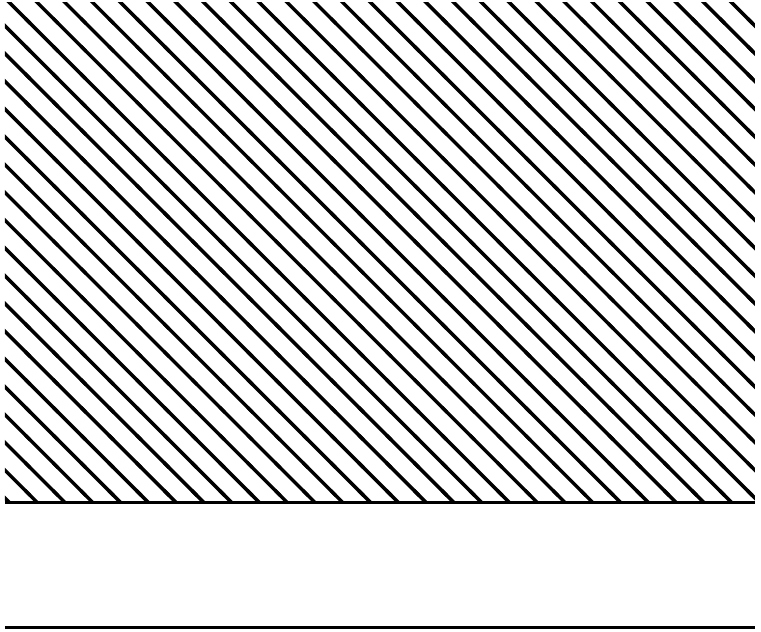}
\end{center}
\caption{The configuration space of the toy example consists of two disconnected parts of different dimensions, a line (bottom) and a half plane (top).}
\label{fig:toyQ}
\end{figure}

Consider, as the configuration space $\Q$, the (disjoint) union of $\Q^{(1)}=\RRR$ and $\Q^{(2)}=\bigl\{(x,y)\in\RRR^2: y\geq 0\bigr\}$; see Figure~\ref{fig:toyQ}. 

We refer to $\Q^{(1)}$ and $\Q^{(2)}$ as the two \emph{sectors} of $\Q$. Wave functions are complex-valued functions on $\Q$; any such function can be specified by specifying $\psi^{(1)}$ and $\psi^{(2)}$, the restrictions of $\psi$ to $\Q^{(1)}$ and $\Q^{(2)}$ (called the sectors of $\psi$). To obtain a Hilbert space $\Hilbert$, we regard $\Q$ as a measure space with measure $\mu$ defined in the obvious way by
\be\label{measuredef}
\mu(S) = \lambda^{(1)}\bigl(S\cap \Q^{(1)}\bigr) + \lambda^{(2)}\bigl(S\cap\Q^{(2)}\bigr)
\ee
for measurable sets $S\subseteq \Q$ with $\lambda^{(n)}$ the Lebesgue measure ($n$-dimensional volume) in $\RRR^n$, and take $\Hilbert=L^2(\Q,\mu)$. That is, the inner product is given by
\be
\scp{\psi}{\phi} = \int\limits_{\Q^{(1)}} \! dx\, \psi^{(1)} (x)^* \, \phi^{(1)}(x) + \int\limits_{\Q^{(2)}} \!\! dx\, dy \, \psi^{(2)}(x,y)^* \, \phi^{(2)}(x,y)\,.
\ee
Equivalently, 
\be
\Hilbert=\Hilbert^{(1)}\oplus \Hilbert^{(2)}
\ee
with
\be
\Hilbert^{(n)}= L^2\bigl(\Q^{(n)},\lambda^{(n)}\bigr) \,.
\ee

The configuration space $\Q$ has a \emph{boundary}
\be
\partial \Q = \bigl\{(x,y)\in\RRR^2: y=0\bigr\}\,.
\ee
That is, $\Q^{(2)}$ has a boundary, while the boundary of $\Q^{(1)}$ is empty. We call any non-boundary point an \emph{interior point}.

\subsection{Interior--Boundary Condition and Hamiltonian}

We now impose a boundary condition on $\psi$. Usual boundary conditions are conditions on the value or a derivative of $\psi$ at a boundary point. However, our boundary condition, which we call an \emph{interior--boundary condition} (IBC), relates the value (or a derivative) of $\psi$ at a boundary point to the value of $\psi$ at an interior point; in this case, the boundary point $(x,0)\in \partial \Q^{(2)}$ gets compared to the point $x\in\Q^{(1)}$, which is an interior point in a different sector. The IBC reads:
\be\label{IBC1}
\psi^{(2)}(x,0) = -\tfrac{2mg}{\hbar^2}\:\psi^{(1)}(x)
\ee
for every $x\in\RRR$. Here, $m>0$ is a mass parameter and $g>0$ a coupling constant. Note that the IBC is a linear condition. We elucidate below how one can arrive at considering this condition. The IBC \eqref{IBC1} goes together with the following Hamiltonian $H$:
\begin{subequations}\label{H1def}
\begin{align}
(H\psi)^{(1)}(x)\: &=-\tfrac{\hbar^2}{2m} \partial^2_x \psi^{(1)}(x) + g  
\,\partial_y \psi^{(2)}(x,0) \label{H1def1} \\
(H\psi)^{(2)}(x,y)\: & =-\tfrac{\hbar^2}{2m} \Bigl(\partial^2_x+\partial_y^2\Bigr) \psi^{(2)}(x,y) \quad \text{for }y>0\,.\label{H1def2}
\end{align} 
\end{subequations}
It consists of the free Schr\"odinger operators and a further term that links $\psi^{(1)}$ to $\psi^{(2)}$. Note that $H$ acts linearly on $\psi=(\psi^{(1)},\psi^{(2)})$. In order to understand the effect of the additional term, and why \eqref{H1def} can be expected to define a unitary time evolution, we need to consider the balance equation for the probability current.

\subsection{Probability Current and Probability Balance}

The well-known probability current vector field associated with the free Schr\"odinger Hamiltonian $-(\hbar^2/2m) \nabla^2$ on $\RRR^n$ has the form
\be\label{jdef}
j = \tfrac{\hbar}{m}\, \Im \bigl[ \psi^* \nabla \psi \bigr]
\ee
and satisfies a continuity equation with the probability density $\rho=|\psi|^2$,
\be\label{continuity}
\partial_t \rho = - \nabla\cdot j\,.
\ee
Generally, it follows from the Schr\"odinger equation $i\hbar\partial_t\psi = H\psi$ that
\be\label{dpsi2dt}
\frac{\partial |\psi(q)|^2}{\partial t} = \tfrac{2}{\hbar}\, \Im \Bigl[ \psi(q)^* (H\psi)(q)\Bigr]
\ee
at any configuration $q$, and for $H=-(\hbar^2/2m) \nabla^2$ the right-hand side becomes that of \eqref{continuity}.

When considering a configuration space with boundary as in our example, the possibility arises of a probability current into the boundary, which can mean a loss of overall probability and thus a breakdown of unitarity. This can be avoided by boundary conditions such as a Dirichlet condition
\be\label{Dirichlet1}
\psi^{(2)}(x,0) =0
\ee
or a Neumann condition
\be\label{Neumann1}
\partial_y \psi^{(2)}(x,0) =0\,.
\ee
Either of these conditions forces the current $j$ to have vanishing normal component at every boundary point, leading to zero current into the boundary. The IBC setup, in contrast, allows nonvanishing current into the boundary while compensating this loss by a gain in probability on a different sector. In fact, the balance equation for the probability density in the first sector reads
\begin{subequations}\label{balance}
\begin{align}
\frac{\partial |\psi^{(1)}|^2}{\partial t} &=\tfrac{2}{\hbar}\, \Im \Bigl[ \psi^{(1)}(x)^* (H\psi)^{(1)}(x) \Bigr]\label{step1}\\
&=  -\partial_x j^{(1)}(x)  + \tfrac{2}{\hbar}\, \Im \Bigl[ \psi^{(1)}(x)^* \,g\,  
\partial_y \psi^{(2)}(x,0)\Bigr]\label{step2}\\
&=  -\partial_x j^{(1)}(x) - \tfrac{2g}{\hbar} \,\Im \Bigl[ 
\tfrac{\hbar^2}{2mg} \psi^{(2)}(x,0)^*
\,\partial_y \psi^{(2)}(x,0) \Bigr] \label{step3}\\
&=  -\partial_x j^{(1)}(x) -  
\tfrac{\hbar}{m}\, \Im \Bigl[ \psi^{(2)}(x,0)^* \; \partial_y \psi^{(2)}(x,0) \Bigr]\label{step4}\\ 
&=  -\partial_x j^{(1)}(x) -   
j_y^{(2)}(x,0)\,, \label{step5}
\end{align}
\end{subequations}
where we have used first 
\eqref{dpsi2dt} in \eqref{step1}, then \eqref{H1def1} in \eqref{step2}, then \eqref{IBC1} in \eqref{step3}, and then \eqref{jdef} in \eqref{step5}, writing $j_y^{(2)}$ for the $y$-component of the 2-vector $j^{(2)}$. The last equation \eqref{step5} means that on $\Q^{(1)}$, $|\psi|^2$ changes due to two factors: transport along $\Q^{(1)}$ as governed by $j^{(1)}$, plus a second term signifying gain or loss in probability that compensates exactly the loss or gain in $\Q^{(2)}$ due to current into the boundary, since the usual continuity equation \eqref{continuity} holds in the interior of $\Q^{(2)}$. In this way, the overall probability $p^{(1)}(t) + p^{(2)}(t)$, with
\be
p^{(1)}(t) :=
\int\limits_{\Q^{(1)}} \! dx \, \bigl| \psi^{(1)}(x,t) \bigr|^2\,, \quad 
p^{(2)}(t) := \int\limits_{\Q^{(2)}} \!\! dx\, dy\, \bigl| \psi^{(2)}(x,y,t) \bigr|^2\,,
\ee
is conserved, while probability may well be exchanged between $\Q^{(1)}$ and $\Q^{(2)}$, so that $p^{(1)}(t)$ and $p^{(2)}(t)$ are not individually conserved. 

Readers may find it useful to visualize the probability flow in terms of Bohmian trajectories \cite{bohmibc}. The Bohmian configuration corresponds to a random point $Q_t$ in configuration space that moves in a way designed to ensure that $Q_t$ has probability distribution $|\psi(t)|^2$ for every $t$. In our example, this distribution entails, when $\psi^{(1)}$ and $\psi^{(2)}$ are both non-zero, that $Q_t$ lies in $\Q^{(1)}$ with probability $p^{(1)}(t)$
and in $\Q^{(2)}$ with probability 
$p^{(2)}(t)$. If in $\Q^{(2)}$, $Q_t$ moves according to Bohm's law of motion
\be\label{Bohm1}
\frac{dQ_t}{dt} = \frac{j^{(2)}(Q_t,t)}{\rho^{(2)}(Q_t,t)}
\ee
until $Q_t$ hits the boundary at $(X,0)$, at which moment the configuration jumps\footnote{An alternative way of looking at the situation, without jumps, arises from identifying $\Q^{(1)}$ with the boundary of $\Q^{(2)}$; this is described under the name ``radical topology'' of $\Q$ in \cite{bohmibc}. Such an identification must be used with care, for example because the measure on $\Q$ is still given by \eqref{measuredef} whereas boundaries usually have measure zero, and because the Laplacian on $\Q^{(1)}$ does not have a $\partial_y^2$ term. If we make this identification, which goes particularly naturally together with the IBC \eqref{IBC1} if $-2mg/\hbar^2=1$, then the Bohmian configuration does not jump, but simply moves along the boundary $\partial \Q^{(2)}=\Q^{(1)}$ after reaching it.} to $X\in\Q^{(1)}$. Once in $\Q^{(1)}$, the configuration moves according to Bohm's law of motion, i.e.,
\be
\frac{dQ_t}{dt} = \frac{j^{(1)}(Q_t,t)}{\rho^{(1)}(Q_t,t)}\,,
\ee
and during any time interval of infinitesimal length $dt$, $Q_t=X_t$ has probability
\be\label{jumprate}
\sigma(X_t,t) \, dt = \frac{\max\{0,
j^{(2)}(X_t,0,t)\}}{\rho^{(1)}(X_t,t)} \, dt\,,
\ee
to jump to the point $(X_t,0)$ on the boundary of $\Q^{(2)}$ and continue from there into the interior of $\Q^{(2)}$ according to \eqref{Bohm1}. From these laws it follows that if $Q_t$ is $|\psi(t)|^2$ distributed for $t=0$, then $Q_t$ is $|\psi(t)|^2$ distributed also for $t>0$ \cite{bohmibc}.

On another matter, it may seem from the defining equations \eqref{H1def} of $H$ that $H$ cannot be Hermitian because there is no Hermitian conjugate to the term $g 
\, \partial_y \psi^{(2)}(x,0)$ in \eqref{H1def1}. However, the conservation of probability just discussed implies that there is no need for such a term; rather, the IBC replaces it.

Let us return once more to the calculation \eqref{balance} to understand how one arrives at the Hamiltonian \eqref{H1def} and the IBC \eqref{IBC1}. Suppose we want \eqref{step5} to hold, i.e., we want an additional term in the balance equation for $\rho^{(1)}$ that compensates the loss or gain of probability in $\Q^{(2)}$ due to current into the boundary. Then we have to have, in the expression for $(H\psi)^{(1)}$, an additional term $T_2$ beyond $-\tfrac{\hbar^2}{2m}\partial_x^2 \psi^{(1)}(x)$, and in order to go from \eqref{step1} to \eqref{step4} we need that
\be\label{needed}
\tfrac{2}{\hbar}\, \Im \Bigl[ \psi^{(1)}(x)^*\, T_2 \Bigr] =  -  
\tfrac{\hbar}{m}\, \Im \Bigl[ \psi^{(2)}(x,0)^* \, \partial_y \psi^{(2)}(x,0) \Bigr]\,.
\ee
This situation suggests that $T_2$ should involve $\psi^{(2)}$. Since $T_2$ needs to be linear in $\psi$, we need another ingredient that will allow us to replace the $\psi^{(1)}(x)^*$ on the left-hand side by a term involving $\psi^{(2)}$, thus leading to an IBC. One possibility is that 
$\psi^{(2)}(x,0)^*$ on the right-hand side is proportional to $\psi^{(1)}(x)^*$ by virtue of the IBC, and that the term $\partial_y \psi^{(2)}(x,0)$ comes from $T_2$, and that leads to the equations we gave above, with an arbitrary choice of the coupling constant $g$. (In fact, we may allow $g$ to be negative or even complex if we replace $g$ by $g^*$ in \eqref{IBC1}. However, this does not really lead to more possibilities, as the resulting time evolution is unitarily equivalent to the one with real coupling constant $|g|$. That is because if $\psi$ satisfies \eqref{IBC1} with $g\to g^*$ and \eqref{H1def} then $\tilde\psi$ with $\tilde\psi^{(2)} = \frac{g}{|g|} \psi^{(2)}$, $\tilde\psi^{(1)}= \psi^{(1)}$ satisfies \eqref{IBC1} and \eqref{H1def} with $g \to |g|$.)

\subsection{Neumann vs.\ Dirichlet Conditions}

Another possibility for fulfilling \eqref{needed} becomes obvious when re-writing
\be
\Im \Bigl[ \psi^{(2)}(x,y)^* \;\partial_y \psi^{(2)}(x,y) \Bigr] \quad \text{as} \quad
-\Im \Bigl[  \partial_y \psi^{(2)}(x,y)^*\;  \psi^{(2)}(x,y) \Bigr]\,,
\ee
viz., that  
$\partial_y\psi^{(2)}(x,0)^*$ is proportional to $\psi^{(1)}(x)^*$, thus leading to a different IBC, while $T_2$ is proportional to $ 
\psi^{(2)}(x,0)$. This leads to the equations
\begin{subequations}
\begin{align}
{\color{purple} 
\partial_y \psi^{(2)}(x,0)}\:& = {\color{purple}+}\tfrac{2mg}{\hbar^2}\psi^{(1)}(x) \quad\quad \text{(IBC)}\label{NeuIBC}\\
(H\psi)^{(1)}(x)\:&=-\tfrac{\hbar^2}{2m} \partial^2_x \psi^{(1)}(x)+ g\,{\color{purple} 
\psi^{(2)}(x,0)} \\
(H\psi)^{(2)}(x,y)\:&=-\tfrac{\hbar^2}{2m} \Bigl(\partial^2_x+\partial_y^2\Bigr) \psi^{(2)}(x,y)\quad \text{for }y>0\,.
\end{align} 
\end{subequations}
instead of \eqref{IBC1} and \eqref{H1def}, which we repeat here for comparison:
\begin{subequations}
\begin{align}
{\color{purple}  
\psi^{(2)}(x,0)}\:& = {\color{purple}-}\tfrac{2mg}{\hbar^2}\psi^{(1)}(x) \quad\quad \text{(IBC)}\label{DirIBC}\\
(H\psi)^{(1)}(x)\:&=-\tfrac{\hbar^2}{2m} \partial^2_x \psi^{(1)}(x) + g\, {\color{purple}  
\partial_y \psi^{(2)}(x,0)} \\
(H\psi)^{(2)}(x,y)\:&=-\tfrac{\hbar^2}{2m} \Bigl(\partial^2_x+\partial_y^2\Bigr) \psi^{(2)}(x,y)\quad \text{for }y>0\,.
\end{align} 
\end{subequations}
That is, while the original IBC \eqref{DirIBC} was of Dirichlet type (in that it specifies the value of $\psi^{(2)}$ on the boundary), the alternative IBC \eqref{NeuIBC} is of Neumann type (in that it specifies the normal derivative of $\psi^{(2)}$ on the boundary). This change is accompanied by a change of the term $T_2$ in the equation for the Hamiltonian and leads to a different time evolution.

Another type of boundary condition often considered besides the Neumann condition
\be
\frac{\partial \psi}{\partial n}\Big|_{\partial \Q}=0
\ee
and the Dirichlet condition
\be
\psi\Big|_{\partial \Q}=0
\ee
is the Robin boundary condition
\be
\alpha \psi + \beta \frac{\partial \psi}{\partial n}\Big|_{\partial \Q}=0
\ee
with constants $\alpha,\beta\in\RRR$. Correspondingly, another possibility for the IBC and the equations for $H$ is
\begin{subequations}
\begin{align}
{\color{purple} 
\bigl( \alpha + \beta \partial_y\bigr)\psi^{(2)}(x,0)}\:& = {\color{purple}+} \tfrac{2mg}{\hbar^2}\psi^{(1)}(x) \quad\quad \text{(IBC)}\label{RobIBC}\\
(H\psi)^{(1)}(x)\:&=-\tfrac{\hbar^2}{2m} \partial^2_x \psi^{(1)}(x) + g\: {\color{purple} 
\bigl(\gamma + \delta \partial_y\bigr) \psi^{(2)}(x,0)} \\
(H\psi)^{(2)}(x,y)\:&=-\tfrac{\hbar^2}{2m} \Bigl(\partial^2_x+\partial_y^2\Bigr) \psi^{(2)}(x,y)\quad \text{for }y>0
\end{align} 
\end{subequations}
with constants $\alpha,\beta,\gamma,\delta\in \RRR$ such that $\alpha\delta-\beta\gamma=-1$.

\subsection{Rigorous, Self-Adjoint Hamiltonian}

Readers that are mathematicians may be interested in the rigorous definition of the Hamiltonian, which we give here for the Dirichlet-type condition \eqref{IBC1}. The domain $\domain$ consists of functions in $\Hilbert$ that satisfy the IBC. More precisely, let
\be\label{domain0def}
\domain_0= H^2(\RRR) \oplus H^2\bigl(\RRR\times[0,\infty)\bigr)\subseteq \Hilbert^{(1)}\oplus \Hilbert^{(2)}\,,
\ee
where $H^2$ denotes the second Sobolev space, and $H^2\bigl(\RRR\times[0,\infty)\bigr)$ contains the restriction of functions in $H^2(\RRR^2)$ to $\RRR\times[0,\infty)$. By the Sobolev imbedding theorem (e.g., \cite[p.~85]{AF}), any element $\psi^{(2)}$ of $H^2(\RRR^2)$ possesses a unique restriction $f\in L^2(\RRR)$ to the subspace $\RRR\times \{0\}$. The IBC \eqref{IBC1}, understood as the condition $f=(-2mg/\hbar^2)\psi^{(1)}$, is thus meaningful for every $\psi\in \domain_0$, and we can define
\be
\domain = \Bigl\{\psi\in\domain_0: f=-\tfrac{2mg}{\hbar^2}\,\psi^{(1)} \Bigr\}\,.
\ee
Likewise, the $y$-derivative of any element of $H^2(\RRR^2)$ lies in $H^1(\RRR^2)$ and possesses, by the Sobolev imbedding theorem, a unique restriction $h\in L^2(\RRR)$ to the subspace $\RRR\times \{0\}$. Thus, the Hamiltonian can be defined on $\domain$ by
\begin{subequations}
\begin{align}
(H\psi)^{(1)}(x)\:&=-\tfrac{\hbar^2}{2m} \partial^2_x \psi^{(1)}(x) + g \, h(x) \\
(H\psi)^{(2)}(x,y)\:&=-\tfrac{\hbar^2}{2m} \Bigl(\partial^2_x+\partial_y^2\Bigr) \psi^{(2)}(x,y)\,,
\end{align} 
\end{subequations}
and one can show:

\begin{thm}
$\domain$ is dense in $\Hilbert$, and $H$ is self-adjoint on the domain $\domain$.
\end{thm}

\section{Particle Creation via IBC}
\label{sec:crea}

We now transfer the IBC approach to a simple model of particle creation and annihilation.

\subsection{Configuration Space and Hilbert Space}

Suppose that $x$-particles can emit and absorb $y$-particles, and consider a single $x$-particle fixed at the origin. For simplicity, we take the $x$- and $y$-particles to be spinless, and we intend to cut off from the Fock space for the $y$-particles any sector with particle number 2 or higher. To this end, we consider only $y$-configurations with 0 or 1 particle, so $\Q=\Q^{(0)}\cup \Q^{(1)}$, where $\Q^{(0)}$ has a single element (the empty configuration $\emptyset$), while $\Q^{(1)}$ is a copy of physical space $\RRR^3$.  Wave functions are again functions $\psi:\Q\to\CCC$, and the Hilbert space is 
\be
\Hilbert=\Hilbert^{(0)}\oplus \Hilbert^{(1)} = \CCC\oplus L^2(\RRR^3)\,,
\ee
which has inner product
\be
\scp{\psi}{\phi} = \psi^{(0)*} \phi^{(0)} + \int\limits_{\Q^{(1)}} \! d^3\vy \: \psi^{(1)}(\vy)^* \, \phi^{(1)}(\vy)\,.
\ee

Writing $\vy\in\Q^{(1)}$, the position of the $y$-particle, in spherical coordinates $(r,\vomega)$ with $0\leq r<\infty$ and $\vomega\in\SSS^2$ (the unit sphere in $\RRR^3$), we can think of $\Q^{(1)}$ as
\be
\Q^{(1)} =[0,\infty)\times \SSS^2
\ee
with Riemannian metric
\be
ds^2 = dr^2 + r^2 \, d\vomega^2
\ee
(with $d\vomega^2$ the 2-dimensional metric on the sphere). The inner product then reads
\be
\scp{\psi}{\phi} = \psi^{(0)*} \phi^{(0)} + \int\limits_0^\infty \! dr \int\limits_{\SSS^2} \! d^2\vomega \: r^2 \, \psi^{(1)}(r,\vomega)^* \, \phi^{(1)}(r,\vomega)\,,
\ee
and the Laplace operator becomes
\be
\Laplace = \partial_r^2+\tfrac{2}{r}\partial_r + \tfrac{1}{r^2} \Laplace_{\vomega}
\ee
with $\Laplace_{\vomega}$ the Laplace operator on the sphere.

\subsection{IBC and Hamiltonian}

The relevant boundary of $\Q^{(1)}$ is the set $\partial \Q^{(1)}=\{r=0\}$, which corresponds to the origin (the location of the $x$-particle) in $\RRR^3$ and is represented by the surface $\{0\}\times \SSS^2$ in spherical coordinates. Probability current into this boundary corresponds to the annihilation of the $y$-particle and current out of the boundary to the creation of a $y$-particle. (In terms of Bohmian trajectories, when a trajectory $Q_t$ in $\Q^{(1)}$ hits $\{r=0\}$, so that the $y$-particle reaches the origin, then $Q_t$ jumps to $\emptyset\in\Q^{(0)}$, so that the $y$-particles gets absorbed by the $x$-particle; conversely, if $Q_t=\emptyset$, then at a random time, governed by a law similar to \eqref{jumprate}, $Q_t$ jumps to $\{r=0\}$ and moves into the interior, $\{r>0\}$, so that a $y$-particle gets emitted by the $x$-particle.)

In this setup, the IBC analogous to \eqref{IBC1} reads: For every $\vomega\in\SSS^2$,
\be\label{IBC9a}
\lim_{r\searrow 0} \bigl(r\psi^{(1)}(r\vomega)  \bigr) 
= -\tfrac{mg}{2\pi\hbar^2}\; \psi^{(0)}\,.
\ee
The Hamiltonian analogous to \eqref{H1def} is
\begin{subequations}\label{Hdef5}
\begin{align}
(H\psi)^{(0)} &= \tfrac{g}{4\pi} \int\limits_{\SSS^2} d^2\vomega\, \lim_{r\searrow 0} \partial_r\Bigl(r\psi^{(1)}(r\vomega)  \Bigr)\label{Hdef5a}\\
(H\psi)^{(1)}(r\vomega) &= -\tfrac{\hbar^2}{2m} \Bigl( \partial_r^2+\tfrac{2}{r}\partial_r + \tfrac{1}{r^2} \Laplace_{\vomega}\Bigr) \psi^{(1)}(r\vomega) \quad \text{for $r>0$}\,.\label{Hdef5b}
\end{align} 
\end{subequations}

It can be shown \cite{ibc2a} that \eqref{Hdef5} defines a self-adjoint operator $H$ on a dense domain $\domain$ in $\Hilbert$ consisting of functions satisfying the IBC \eqref{IBC9a}. (It turns out that elements of $\domain$ satisfy a stronger version of \eqref{IBC9a} that has the limit ${r\to 0}$ replaced by the limit ${r\vomega\to \vzero}$; that is, the stronger version does not demand that the limit be taken \emph{in the radial direction, keeping $\vomega$ constant}, but allows any way of approaching the boundary surface, even without a limiting value for $\vomega$.)

\subsection{Remarks}

\begin{enumerate}
\setcounter{enumi}{\thetumulkaremarks}
\item \textit{$1/r$ asymptotics.}\label{rem:1/r}
As a consequence of the IBC \eqref{IBC9a}, whenever $\psi^{(0)}$ is nonzero then $\psi^{(1)}$ diverges at $\{r=0\}$ like $1/r$. This behavior is to be expected, for a reason that is perhaps most easily appreciated by means of the following simple approximation: Consider an ensemble of systems in which the $y$-particle moves at unit speed towards the origin and gets annihilated when it reaches the origin; that is, the motion follows the equation of motion
\be\label{toyODE}
\frac{d\vy}{dt}=-\frac{\vy}{|\vy|}\,.
\ee
Suppose first that the ensemble density is uniform over a spherical shell of radius $r$ and thickness $dr$; as the members of the ensemble move inwards, the density increases like $1/r^2$ since the area of a sphere is proportional to $r^2$ and the thickness $dr$ remains constant. As a consequence, the stationary density for the particle motion \eqref{toyODE} diverges at the origin like $1/r^2$. It thus comes as no surprise that $|\psi|^2$ on $\Q^{(1)}$ should diverge at the origin like $1/r^2$, and so $|\psi|$ should diverge like $1/r$. Conversely, the $1/r$ divergence of $\psi^{(1)}$ at the boundary $\{r=0\}$ makes the $r$ factor in the IBC \eqref{IBC9a} necessary, as $\lim_{r\searrow 0} \psi^{(1)}(r\vomega)$ (without the $r$ factor) does not exist.

\item \textit{$H_{IBC}$ is not a perturbation of $H_{\free}$.}
We note that $H_{IBC}$ cannot be decomposed into a sum of two self-adjoint operators $H_{\free}+H_{\mathrm{interaction}}$. As a consequence, $H_{IBC}$ cannot be found when studying Hamiltonians of the form $H_{\free}+H_{\mathrm{interaction}}$. That is because the domain $\domain_{IBC}$ of $H_{IBC}$ is different from the free domain $\domain_{\free}$; specifically, functions in $\domain_{IBC}$ diverge like $1/r$ at $\{r=0\}$, while functions in the free domain (the second Sobolev space) stay bounded at $\{r=0\}$ and thus yield $\lim_{r\searrow 0} \bigl(r\psi^{(1)}(r\vomega) \bigr)=0$. The Laplacian is not self-adjoint on $\domain_{IBC}$ (i.e., does not conserve probability) because it allows a nonzero flux of probability into the boundary $\{r=0\}$, while the additional term in $H_{IBC}$ compensates that flux by adding it to $\conf^{(0)}$.

\item \textit{Comparison to a known boundary condition.}
Boundary conditions at $\{r=0\}$ have been used before; in particular, Bethe and Peierls \cite{BP35} introduced the boundary condition
\be\label{BP}
\lim_{r\searrow 0} \Bigl[ \partial_r \bigl(r\psi(r\vomega)\bigr) + \alpha r\psi(r\vomega) \Bigr] = 0 \quad \forall \vomega \in\SSS^2
\ee
with given constant $\alpha\in\RRR$ and wave function $\psi:\RRR^3\to\CCC$, for the purpose of making precise what it means to have on $\Hilbert=L^2(\RRR^3,\CCC)$ a Schr\"odinger equation with a Dirac $\delta$ function as the potential,
\be
H = -\frac{\hbar^2}{2m}\nabla^2 + g\,\delta^3(\vx)\,,
\ee
see \cite{AGHKH88} for more detail. Note that \eqref{BP} leads to zero current into $\{r=0\}$, as that current is
\begin{subequations}
\begin{align}
J_0 
&\hspace{1mm}=\hspace{1mm} -\lim_{r\searrow 0} \int\limits_{\SSS^2}\!d^2\vomega\: r^2\, j_r(r\vomega) \\
&\hspace{1mm}=\hspace{1mm} -\lim_{r\searrow 0} \int\limits_{\SSS^2}\!d^2\vomega\: r^2\, \tfrac{\hbar}{m} \, \Im \bigl[ \psi(r\vomega)^*\, \partial_r \psi(r\vomega) \bigr]\\
&\hspace{1mm}=\hspace{1mm} -\lim_{r\searrow 0} \int\limits_{\SSS^2}\!d^2\vomega\: r\, \tfrac{\hbar}{m} \, \Im \bigl[ \psi(r\vomega)^*\, \partial_r \bigl(r\psi(r\vomega) \bigr) \bigr]\\
&\stackrel{\eqref{BP}}{=} \lim_{r\searrow 0} \int\limits_{\SSS^2}\!d^2\vomega\: r\,\tfrac{\hbar}{m} \, \Im \bigl[ \psi(r\vomega)^*\, \alpha r\psi(r\vomega) \bigr]\\
&\hspace{1mm}=\hspace{1mm}0\,.
\end{align}
\end{subequations}
In contrast, the IBC \eqref{IBC9a}, which we may re-write in the form
\be\label{IBC9b}
\lim_{r\searrow 0} \, r\psi(r\vomega)
= \alpha \, \psi(\emptyset) \quad \forall \vomega\in\SSS^2
\ee
with suitable constant $\alpha\in\RRR$, leads to nonzero current into $\{r=0\}$. Moreover, the IBC \eqref{IBC9b} involves two sectors of $\psi$, while the Bethe--Peierls boundary condition \eqref{BP} involves only one.
\end{enumerate}
\setcounter{tumulkaremarks}{\theenumi}

\section{IBC in QFT}
\label{sec:qft}

The application of interior--boundary conditions in quantum field theory is based on the particle-position representation, in which a QFT becomes a kind of quantum mechanics with particle creation and annihilation. 

In relativistic QFT, there are issues with the particle-position representation, but they do not seem fatal for the IBC approach: 
(i)~Some QFTs appear to lead to an infinite number of particles (e.g., \cite{DDMS10}); a configuration space for an infinite number of particles will be more difficult, but not impossible, to deal with.
(ii)~Photons are believed not to have a good position representation (e.g., \cite{BB}). However, photon wave functions are believed to be mathematically equivalent to (complexified) classical Maxwell fields \cite{BB}, and that may be good enough for IBCs. 
(iii)~The construction of the configuration space is based on a choice of hypersurface in space-time $\sM$; however, the use of multi-time wave functions \cite{PT14} would avoid such a choice, as such wave functions are defined on (the spacelike subset of) $\cup_{n=0}^\infty \sM^n$.

\begin{figure}[h]
\begin{center}
\includegraphics[width=.6 \textwidth]{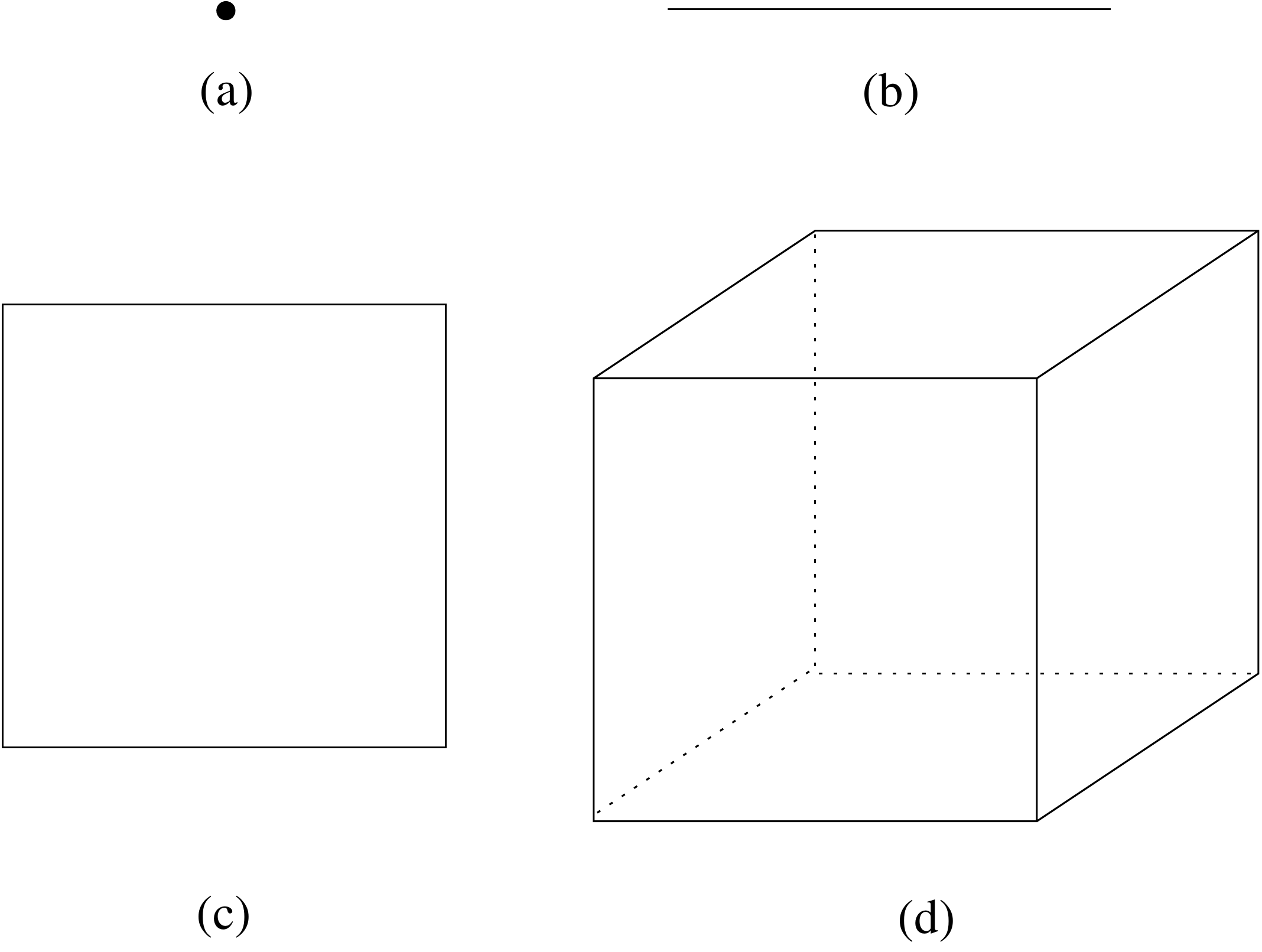}
\end{center}
\caption{The configuration space considered in \eqref{Qdef1} is the disjoint union of $n$-particle configuration spaces (shown here for 1 rather than 3 space dimensions); parts (a) through (d) show the 0-particle through 3-particle sector of the configuration space $\Q$.}
\label{fig:Fock}
\end{figure}

We focus here on non-relativistic models, for which the relevant Hilbert spaces are bosonic or fermionic Fock spaces $\Fock^{\pm}$, or tensor products of such spaces. 
The corresponding configuration space contains configurations of any number of particles (see Figure~\ref{fig:Fock}),
\be\label{Qdef1}
\Q=\bigcup_{n=0}^\infty \Q^{(n)} = \bigcup_{n=0}^\infty (\RRR^3)^n\,.
\ee
In fact, for spinless particles, $\Fock^{\pm}$ consists of those functions $\psi:\Q\to\CCC$ that are (anti-)symmetric on every sector $\Q^{(n)}$ and that are square-integrable in the sense
\be
\sum_{n=0}^\infty\: \int\limits_{\RRR^{3n}} \! d^{3n}q\: |\psi(q)|^2 < \infty \,.
\ee

It is sometimes preferable to consider a slightly different configuration space and remove the collision configurations (i.e., those with two or more particles at the same location),
\be\label{Qndef2}
\Q^{(n)} = \Bigl\{ (\vx_1,\ldots,\vx_n)\in(\RRR^3)^n: \vx_i\neq \vx_j \text{ for }i\neq j \Bigr\}\,.
\ee
Alternatively, it is sometimes desirable to consider \emph{unordered} configurations (e.g., \cite{fermionic}),
\be\label{Qndef3}
\Q^{(n)} = \Bigl\{ q\subset \RRR^3: \# q = n \Bigr\}\,.
\ee

\subsection{Model QFT}
\label{sec:Model1}

Suppose again that $x$-particles can emit and absorb $y$-particles. In \cite{ibc2b} we study a model QFT adapted from \cite[p.~339]{Schw61}, \cite{Nel64}, and the Lee model \cite{Lee54}, starting out from the following, UV divergent expression for the Hamiltonian:
\begin{align}
H_\orig &= \tfrac{\hbar^2}{2m_{x}}\int d^3\vq \, \nabla a_{x}^\dagger(\vq) \, \nabla a_{x}(\vq)\nonumber\\
&\quad+ \tfrac{\hbar^2}{2m_{y}}\int d^3\vq \, \nabla a_{y}^\dagger(\vq) \, \nabla a_{y}(\vq) 
+ E_0 \int d^3\vq\, a_y^\dagger(\vq)\, a_y(\vq)\nonumber\\
&\quad+ g \int d^3\vq\, a_x^\dagger(\vq) \, \bigl(a_y(\vq)+a^\dagger_y(\vq) \bigr)\, a_x(\vq)\,.\label{Horigdef0a}
\end{align}
Here, $^\dagger$ denotes the adjoint operator, and $a(\vq)$ and $a^\dagger(\vq)$ are the annihilation and creation operators for either an $x$- or a $y$-particle at location $\vq$ in position space; $g>0$ a coupling constant, and $E_0\geq 0$ the amount of energy required to create a $y$-particle. The Hilbert space is
\be
\Hilbert = \Fock_x^- \otimes \Fock_y^+\,,
\ee
and the configuration space is
\be
\conf=\bigcup_{m,n=0}^\infty (\RRR^3_x)^m \times (\RRR^3_y)^n\,.
\ee

We consider here a simplified version of this model, in which we allow only a single $x$-particle (but any number of $y$-particles), and this $x$-particle cannot move but is fixed at the origin $\vzero\in\RRR^3$; such models are sometimes connected with the name of van Hove \cite{vH52,Der03}. The Hilbert space of this model is $\Hilbert=\Fock_y^+$, and its configuration space is $\Q=\Q_y = \bigcup_{n=0}^\infty (\RRR^3_y)^n$. For a point $y=(\vy_1,\ldots,\vy_n)$ in configuration space $\Q$, we will often use the notation $y^n$ to convey that this configuration has $n$ $y$-particles. The original Hamiltonian \eqref{Horigdef0a} simplifies in this model to
\begin{align}
H_\orig &=  \tfrac{\hbar^2}{2m_{y}}\int d^3\vq \, \nabla a_{y}^\dagger(\vq) \, \nabla a_{y}(\vq) 
+ E_0 \int d^3\vq\, a_y^\dagger(\vq)\, a_y(\vq)\nonumber\\
&\quad+ g \, \bigl(a_y(\vzero)+a^\dagger_y(\vzero) \bigr)\,.\label{Horigdef1a}
\end{align}
In the particle-position representation, in which elements $\psi$ of $\Hilbert$ are regarded as functions $\psi:\conf\to\CCC$, this reads
\begin{align}
(H_\orig \psi)(y^n) &= 
-\tfrac{\hbar^2}{2m_y} \sum_{j=1}^n \nabla_{\vy_j}^2 \psi(y^n) + nE_0 \psi(y^n) \nonumber\\
&\quad +\: g \sqrt{n+1} \,\psi\bigl(y^n,\vzero\bigr)\nonumber\\
&\quad +\: \frac{g}{\sqrt{n}} \sum_{j=1}^n  \delta^3(\vy_j)\,\psi\bigl(y^n\setminus \vy_j\bigr)\,,\label{Horigdef1}
\end{align}
with the notation $y^n\setminus \vy_j$ meaning $(\vy_1,\ldots,\vy_{j-1},\vy_{j+1},\ldots,\vy_n)$ (leaving out $\vy_j$). $H_\orig$ is UV divergent because the wave function of the newly created $y$-particle, $\delta^3(\vy)$, does not lie in $L^2(\RRR^3)$ (or, has infinite energy).

To obtain a well-defined Hamiltonian, a standard approach is to ``smear out'' the $x$-particle at $\vzero$ with ``charge distribution'' $\varphi(\cdot)$, where the ``cut-off function'' $\varphi$ lies in $L^2(\RRR^3,\CCC)$:
\begin{align}
(H_\cutoff \psi)(y^n) &= 
-\tfrac{\hbar^2}{2m_y} \sum_{j=1}^n \nabla_{\vy_j}^2 \psi(y^n) + nE_0 \psi(y^n) \nonumber\\
&\quad +\: g \sqrt{n+1}  \int_{\RRR^3} d^3\vy\, \varphi(\vy)^*\, \psi\bigl(y^n,\vy\bigr) \nonumber\\
&\quad +\: \frac{g}{\sqrt{n}}  \sum_{j=1}^n \varphi(\vy_j) \,\psi\bigl(y^n\setminus \vy_j\bigr)\,. \label{Hcutoffdef}
\end{align}

\subsection{IBC Approach}
\label{sec:Model1IBC}

What the IBC approach yields for this model is just the extension of the equations of Section~\ref{sec:crea} to an unbounded number of $y$-particles. The relevant boundary of $\Q$ consists of those configurations $y^n$ for which a $y$-particle collides with the $x$-particle, i.e., $\vy_j=\vzero$ for some $j\leq n$; the related interior configuration is obtained by removing $\vy_j$ (and all other $y$-particles at $\vzero$, if any). The Dirichlet-type IBC reads as follows:
For every $y^n\in(\RRR^3\setminus\{\vzero\})^n$ and every $j\leq n$,
\be\label{IBC2}
\lim_{\vy_j\to \vzero} \,|\vy_j| \,\psi(y^n)
= -\tfrac{mg}{2\pi\hbar^2\sqrt{n}}\, \psi(y^n\setminus \vy_j)
\ee
with $m=m_y$. The corresponding Hamiltonian is 
\begin{align}
&(H_{IBC}\psi)(y^n) 
=-\tfrac{\hbar^2}{2m} \sum_{j=1}^{n} \nabla^2_{\vy_j}\psi+ nE_0 \psi \nonumber\\
& + \frac{g\sqrt{n+1}}{4\pi}\int\limits_{\SSS^2} \!\! d^2\vomega \, \lim_{r\searrow 0} \partial_r \Bigl[ r \psi\bigl(y^n,r\vomega \bigr) \Bigr]\nonumber\\
& +\: \frac{g}{\sqrt{n}} \sum_{j=1}^n  \delta^3(\vy_j)\,\psi\bigl(y^n\setminus \vy_j\bigr)\,.\label{Hdef2a}
\end{align}

\begin{thm} {\rm \cite{ibc2a}} \label{thm:Model1}
On a certain dense subspace $\domain_{IBC}$ of $\Hilbert$, the elements of which satisfy the IBC \eqref{IBC2}, the operator $H_{IBC}$ given by \eqref{Hdef2a} 
is well-defined and self-adjoint.
\end{thm}

It may seem that $H_{IBC}$ as in \eqref{Hdef2a} should have the same UV problem as $H_{\orig}$ in \eqref{Horigdef1}; after all, we said that the problem with $H_{\orig}$ is caused by the Dirac $\delta$ function, and the last line of \eqref{Hdef2a} coincides with that of \eqref{Horigdef1}, and in particular contains the same $\delta$ function. And yet, $H_{IBC}$ is well defined and $H_{\orig}$ is not! Here is why.  As in the model of Section~\ref{sec:crea} (see Remark~\ref{rem:1/r}), $\psi$ grows like $1/r=1/|\vy_j|$ as $\vy_j\to\vzero$ due to the IBC \eqref{IBC2}, and as readers may recall from classical electrostatics, where $1/r$ comes up as the Coulomb potential, the Laplacian of $1/r$ (which equals the charge density, according to the Poisson equation of electrostatics) is $-4\pi\delta^3(\vx)$. As a consequence, the Laplacian in the first row of \eqref{Hdef2a} contributes a $\delta$ function, which then gets exactly canceled by the $\delta$ function in the last row of \eqref{Hdef2a}. That is how $H_{IBC}\psi$ can manage to be a square-integrable function on $\Q$.

The fact that the last line of \eqref{Hdef2a} coincides with that of \eqref{Horigdef1}, that is, that both equations have the same term for particle creation, underlines the parallel between $H_{IBC}$ and $H_{\orig}$ and suggests that $H_{IBC}$ may be regarded as a precise interpretation of the formal expression \eqref{Horigdef1} for $H_{\orig}$.

At this point, readers may wonder why the formula \eqref{Hdef5b} for the Hamiltonian in the simpler creation model did not contain a $\delta$ function. The reason is merely a matter of notation, as \eqref{Hdef5} can be equivalently rewritten as
\begin{subequations}\label{Hdef6}
\begin{align}
(H\psi)^{(0)} &= \tfrac{g}{4\pi} \int\limits_{\SSS^2} d^2\vomega\, \lim_{r\searrow 0} \partial_r \Bigl(r\psi^{(1)}(r\vomega)  \Bigr)\label{Hdef6a}\\
(H\psi)^{(1)}(\vy) &= -\tfrac{\hbar^2}{2m} \nabla_{\vy}^2 \psi^{(1)}(\vy) + g \, \delta^3(\vy) \, \psi^{(0)} \,.\label{Hdef6b}
\end{align} 
\end{subequations}
In \eqref{Hdef5b}, we explicitly assumed $r>0$, thus stating the action of $H$ only \emph{away from the origin}, so that the term involving $\delta^3(\vy)$ does not show up. In \eqref{Hdef2a} and \eqref{Hdef6b}, in contrast, we did not exclude the origin because we wanted to make the $\delta$ behavior explicit.

\subsection{Remarks}

\begin{enumerate}
\setcounter{enumi}{\thetumulkaremarks}
\item \textit{Positive Hamiltonian.} For $E_0>0$, it can be shown \cite{ibc2a} that $H_{IBC}$ as in \eqref{Hdef2a} is a \emph{positive} operator. 
\item \textit{Ground state.} It can be shown further \cite{ibc2a} for $E_0>0$ that $H_{IBC}$ possesses a non-degenerate ground state $\psi_{\min}$, which is
\be
\psi_{\min}(\vy_1,\ldots,\vy_n) =\mathcal{N}\frac{(-gm)^n}{(2\pi\hbar^2)^n\sqrt{n!}}\prod_{j=1}^n\frac{e^{-\sqrt{2mE_0}|\vy_j|/\hbar}}{|\vy_j|} 
\ee
with normalization constant $\mathcal{N}$ and eigenvalue
\be
E_{\min}=\frac{g^2m\sqrt{2mE_0}}{2\pi\hbar^3}\,.
\ee
That is, the $x$-particle is dressed with a cloud of $y$-particles.

\item \textit{Effective Yukawa potential between $x$-particles.} 
To compute the effective interaction between $x$-particles by exchange of $y$-particles, 
consider two $x$-particles fixed at $\vx_1=(0,0,0)$ and $\vx_2=(R,0,0)$;
two IBCs, one at $\vx_1$ and one at $\vx_2$; and
two creation and annihilation terms in $H_{IBC}$.
For $E_0>0$, the ground state is
\be
\psi_{\min}(\vy_1,\ldots,\vy_n)=c_n \prod_{j=1}^n \sum_{i=1}^2 \frac{e^{-\sqrt{2mE_0}|\vy_j-\vx_i|/\hbar}}{|\vy_j-\vx_i|} 
\ee
with suitable factors $c_n$ and eigenvalue
\be\label{Yukawa}
E_{\min}=\frac{g^2m}{\pi\hbar^2}\biggl( \frac{\sqrt{2mE_0}}{\hbar}-\frac{e^{-\sqrt{2mE_0}R/\hbar}}{R} \biggr)\,.
\ee
As a consequence, for any two locations $\vx_1$ and $\vx_2$, the ground state energy of the $y$-particles, given the $x$-particles at $\vx_1$ and $\vx_2$, is given by \eqref{Yukawa} with $R=|\vx_1-\vx_2|$. Regarding this energy function of $\vx_1$ and $\vx_2$ as an effective potential for the $x$-particles (which is appropriate when the $x$-particles move slowly, see, e.g., \cite{T03}), we see that $x$-particles effectively interact through an attractive Yukawa potential, $V(R)=\text{const.}-e^{-\alpha R}/R$.

\item \textit{Comparison to renormalization procedure.}
Returning to the scenario with a single $x$-particle fixed at the origin, consider $H_{\cutoff}=H_\varphi$ as in \eqref{Hcutoffdef} with cut-off function $\varphi$ and take the limit $\varphi\to \delta^3$. It is known \cite{Der03}
that, if $E_0>0$, there exist constants $E_\varphi\to \infty$ and a self-adjoint operator $H_\infty$ such that 
\be
H_\varphi-E_\varphi\to H_\infty\,.
\ee
It can be shown \cite{ibc2a} that
\be
H_\infty=H_{IBC}+ \text{const.}\,.
\ee
\end{enumerate}
\setcounter{tumulkaremarks}{\theenumi}

\bigskip

\noindent \textit{Note added.} After completion of this article we have become aware that Equations (35) and (36) were already considered in \cite{Tho84} and \cite{Yaf92}.

\bigskip

\noindent\textit{Acknowledgments.} 
We are grateful to our collaborators Jonas Lampart and Julian Schmidt for their help in developing this approach. We also thank Klaus Fredenhagen, Sheldon Goldstein, Harald Grosse,  Stefan Keppeler, and Michael Kiessling for helpful discussions.
R.T.\ was supported in part by grant no.\ 37433 from the John Templeton Foundation.

\end{document}